\documentclass{elsart}
\usepackage{amssymb}
\usepackage{amsfonts}
\usepackage{amsmath}

\begin{document}

\begin{frontmatter}

\title{Deriving thermal lattice-Boltzmann models from the continuous Boltzmann equation:  theoretical aspects}

\author{P. C. Philippi},
\ead{philippi@lmpt.ufsc.br}
\collab{L. A. Hegele Jr., L. O. E. dos Santos, R. Surmas}
\address{Mechanical Engineering Department.
Federal University of Santa Catarina. 88040-900 Florian\'opolis. SC. Brazil.}

\begin{abstract}
The particles model, the collision model, the polynomial development used for the equilibrium distribution, the time discretization and the velocity discretization are factors that let the lattice Boltzmann framework (LBM) far away from its conceptual support: the continuous Boltzmann equation (BE). Most collision models are based on the BGK, single parameter, relaxation-term leading to constant Prandtl numbers. The polynomial expansion used for the equilibrium distribution introduces an upper-bound in the local macroscopic speed.  Most widely used time discretization procedures give an explicit numerical scheme with second-order time step errors.  In thermal problems, quadrature did not succeed in giving discrete velocity sets able to generate multi-speed regular lattices. All these problems, greatly, difficult the numerical simulation of LBM based algorithms.  In present work, the systematic derivation of lattice-Boltzmann models from the continuous Boltzmann equation is discussed. The collision term in the linearized Boltzmann equation is modeled by expanding the distribution function in Hermite tensors. Thermohydrodynamic macroscopic equations are correctly retrieved with a second-order model. Velocity discretization is the most critical step in establishing regular-lattices framework. In the quadrature process, it is shown that the integrating variable has an important role in defining the equilibrium distribution and the lattice-Boltzmann model, leading, alternatively, to temperature dependent velocities (TDV) and to temperature dependent weights (TDW) lattice-Boltzmann models. 
\end{abstract}


\end{frontmatter}

\section*{Introduction}

Following Lallemand \& Luo, \cite{lallemandluo}, the presently known
lattice-Boltzmann equation (LBE) has not been able to handle realistic
thermal (and fully compressible) fluids with satisfaction. Simulation of
thermal lattice-Boltzmann equation is hampered by numerical instabilities
when the local velocity increases. Readers are referred to this work for an
excellent review of known works on thermal and compressible
lattice-Boltzmann schemes.\newline
Rigorously, fluid flow is never isothermal. Consider, for instance, a
channel Poiseuille flow between two adiabatic solid surfaces. Due to the
viscous conversion of mechanical in thermal energy, temperature will vary,
attaining a minimum at the channel symmetry axis, where the local speed is a
maximum. This temperature variation can be very small, but increases with
the local average speed $u$ and with the fluid viscosity. If it is agreed
that the temperature in a given site is related to the expected value E of
the lattice-particles fluctuation kinetic energy, this temperature variation
is drafted in LB athermal simulation, since E varies from site to site in
accordance with the local macroscopic speed, $u$, attaining a minimum where $%
u$ is maximum. However, in athermal lattice-Boltzmann models, this thermal
spatial non-equilibrium is not properly compensated by heat flow because
athermal models were not conceived for correctly describing energy transfer.
In this manner, since temperature gradients cannot be avoided in athermal LB
simulation, in contrast with classical CFD isothermal simulation, they may
become sources of numerical instability.\newline
In conclusion, lattice-Boltzmann athermal\textit{\ }equation, actually,
deals with thermal problems and \textit{thermal} and \textit{athermal}
lattice-Boltzmann models will be here considered using a single approach.%
\newline
There are several features that let the lattice Boltzmann, regular-lattice
based, framework far away from what it would desirable to be its starting
point: the continuous Boltzmann equation. These features include the
particles model, the collision model, the polynomial development used for
the equilibrium distribution, the time discretization and the velocity
discretization. Some of these main features are discussed in the following.%
\newline
\textbf{Collision model. }More widely used lattice-Boltzmann collision
models are based on a Bhatnagar, Gross and Krook (BGK) relaxation term, \cite%
{bgk}, firstly introduced in the lattice-Boltzmann framework by Qian \textit{%
et al.,} \cite{qian},\textit{\ }and Chen\textit{\ et al., }\cite{chen}%
\textit{.} Thermal lattice-Boltzmann schemes based on the BGK collision
model use an increased number of discrete velocities and/or include higher
order non-linear terms in the equilibrium distribution function (\cite%
{alexander}, \cite{mcnamara}, \cite{ychen}), temperature dependent weights, 
\cite{alexander}, and temperature dependent velocities, \cite{vahala}. BGK
single relaxation time collision term restricts the models to constant
Prandtl number. The correct description of fluids and fluid flow requires
multiple relaxation time models (MRT). A two-parameters model was introduced
by He \textit{et al}., \cite{hechendoolen}, using two sets of distributions
for the particles number density and for the thermodynamic internal energy,
coupled through a viscous dissipation term. Full MRT models were firstly
introduced in the LBE framework by d'Humi\`{e}res, \cite{dhumieres}, derived
from the moments method, by making the moments and the velocity spaces
isomorphic, \cite{dhumieresetal}. The main criticism to the moments method
is that the highest order of the kinetic moments possible to be correctly
described with the LBE equation is limited by the number of lattice
velocities, \cite{heluo}, and high-order kinetic moments are not correctly
described when all the b-moments in a b-discrete velocities set are
considered. In currently produced works dealing with applications of the
moments \ method, e.g. \cite{lallemandluo}, the main worry is \textit{%
numerical stability} and \textit{not} the description of non-isothermal,
multicomponent or immiscible fluids flows, which, effectively, require
additional relaxation parameters with respect to BGK models. Dispersion
equations are used as constraints for the adjustable parameters related to
the short wave-length non-hydrodynamic moments and numerical stability is
assured by buffering these higher frequency moments. This artificial
shielding can be too dangerous for the complex flow structures that
naturally appear when the Reynolds number increases. A correct description
of the growing of these flow structures with the Reynolds number requires to
increase the lattice dimensionality.\newline
\textbf{Time discretization. }Most lattice-Boltzmann simulations are based
on an explicit numerical scheme, with second order time-step, $\delta $,
errors. Lattice BGK models, have been simulated with implicit numerical
schemes, \cite{krafk}, \cite{sanka}, or LBE modified explicit numerical
schemes, \cite{hechendoolen}, with third order time step, O($\delta ^{3}$),
errors. In spite of the fact that, in athermal models, this truncation error
can be totally absorbed into the physical viscous term, in thermal models
errors O($\delta ^{2}$) seriously affects the viscous heat dissipation term, 
\cite{hechendoolen}.\newline
\textbf{Velocity discretization. }Lattice-Boltzmann method is based on a
finite set of discrete velocities $\vec{c}_{i}$ and weights $\omega _{i}$,
judiciously chosen so as to ensure isotropy for the even-parity rank
velocity tensors and, consequently, for the fluid transfer properties. He \&
Luo,\cite{heluo}, have directly derived some widely used lattices (D2Q9,
D2Q6, D2Q7, D3Q27) from the continuous Boltzmann equation by discretization
of the velocity space, using the Gauss-Hermite and Gauss -Radau quadrature
of the Maxwellian $\vec{u}$-polynomial expansions. Unhappily, excluding the
above mentioned lattices, the discrete velocity sets obtained by quadrature
do not generate regular lattices. In this sense, Pavlo \textit{et al.}, \cite%
{vahala}, proposed a temperature dependent velocity model based on an
octagonal lattice which is not space-filling but ensures the isotropy of 6$%
^{th}$- rank velocity tensors.\newline
In this work, we present an attempt for deriving the lattice-Boltzmann
equation, from the continuous Boltzmann equation trying to combine the
following main features: multiple relaxation-times, O($\delta ^{3}$) time
step errors and thermodynamic consistency in non-isothermal flow.\newline
In contrast with the moments method, in present conception higher Reynolds
flows (and non-isothermal flows) require to increase the number of the
lattice discrete velocities and to increase the accuracy of the LBE equation
with respect to its continuous counterpart.\newline
The collision term $\Omega $ in the linearized Boltzmann equation is modeled
by expanding the distribution function $f$ \ in Hermite polynomial tensors $%
\Psi _{\theta }$, which forms an orthogonal basis in the Hilbert space $%
\mathcal{H}$ generated by $h:\mathcal{C}^{D}\rightarrow \mathcal{R}$, $D$
being the dimension of the velocity space. Considering that each term $%
\mathcal{L}$ ($\Psi _{\theta }$ ) is, itself, an element of $\mathcal{H}$,
this term is expanded as a linear combination of the same order-$\theta $
Hermite tensors through 2$\theta $-order relaxation tensors. Isotropy
properties are used to reduce these tensors. The infinite series $f^{eq}%
\mathcal{L}$ ($\phi $) is not truncated. Instead, after a chosen tensor
order $N$, a Gross-Jackson procedure is used, \cite{cercignani}, and the
relaxation tensors are diagonalized.\newline
It is shown that d\'Humi\`{e}res moment equations \cite{dhumieresetal} are
particular discrete forms of the derived model when the diagonalization
constant is considered to be zero.\newline
By performing a Chapman-Enskog analysis of the derived continuous model it
is shown that the thermo-hydrodynamic macroscopic equations are correctly
retrieved with a second-order model. Third-order models are only required
for describing third-order coupling (Soret and Dufour effects) in
multi-component systems (\cite{philippi}).\newline
The derived kinetic model of the continuous Boltzmann equation is then
discretized.\newline
It is shown that an explicit numerical scheme with $O$($\delta ^{3}$) time
step errors can be derived, using He \textit{at al. }procedure\textit{, } 
\cite{hechendoolen}.\newline
Velocity discretization is the most critical step in deriving
lattice-Boltzmann equations.\newline
For each $N$, the equilibrium distribution is taken as an $n^{th}$-degree
Hermite development of the Maxwell-Boltzmann (MB) equilibrium distribution,
in accordance with the constraints that are imposed by the physical problem.%
\newline
Although, as it was shown in \cite{heluo}, velocity discretization of the
most widely known lattices can be achieved by Gauss-Hermite and related
quadratures, quadrature schemes did not succeed in generating multi-speed
regular lattices, suitable for thermal problems, placing an, still, open
question in the LBM framework. In present paper, it is shown that the
integrating variable has an important role in defining the equilibrium
distribution and the lattice-Boltzmann model: a) discretization based on the
particles velocity $\vec{c}$, giving a set of discrete, constant, particle
velocities $\vec{c_{i}}$, leads to temperature dependent weights $\omega
_{i} $ (TDW), b) discretization based on the, temperature dependent,
dimensionless velocity $\vec{\mathcal{C}}$ \ gives a set of temperature
dependent particle velocities $\vec{c_{i}}$\ \ (TDV).\newline
In this context, it is shown that the thermal part $g_{i}^{eq}$ in He 
\textit{et al.} two-distributions model, \textit{\ }\cite{hechendoolen}, can
be formally retrieved from TDW models, as an $\vec{u}$-polynomial
approximation with errors $O\left( u\Theta \right) $ where $\Theta $ is the
temperature deviation.\newline
Although a more complete theoretical analysis is still necessary, the
consideration of TDV models appears to be suitable for thermal problems.
However, in this case, the particles allocation rules, required for the
local conservation of mass, momentum and energy, make the particles
number-density, $n$, to be temperature dependent and the implicit
temperature dependence of $n$ is difficult to manage, when performing a
Chapman-Enskog analysis of the kinetic models. Finding the macroscopic
behavior of these models is, still, in progress.\newline
A simulation scheme suitable for non-isothermal problems is presented for
the TDV model.

\section{ Boltzmann equation in the continuous velocity space}

\subsection{Development of the distribution function in Hermite polynomials}

The Maxwell-Boltzmann equilibrium distribution, \cite{cercignani}, can be
written as an infinite series of Hermite polynomial tensors $\Psi _{\theta
,(r_{_{\theta }})}$, \cite{grad}, 
\begin{equation}
f^{eq}=\frac{e^{-\mathcal{C}^{2}}}{\pi ^{D/2}}\left( \frac{m}{2kT}\right)
^{D/2}\sum_{\theta }a_{\theta ,(r_{_{\theta }})}^{eq}\Psi _{\theta
,(r_{_{\theta }})},  \label{serieseq}
\end{equation}%
where $(r_{\theta })$ is a sequence of indexes $r_{1},r_{2},...r_{\theta }$
and repeated index means summation, $\Psi _{0}=1,\Psi _{1,\alpha }=2\mathcal{%
C}_{\alpha }$, $\Psi _{2,\alpha \beta }=2(\mathcal{C}_{\alpha }\mathcal{C}%
_{\beta }-\frac{1}{2}\delta _{\alpha \beta })$, $\Psi _{3,\alpha \beta
\gamma }=\frac{4}{3}(\mathcal{C}_{\alpha }\mathcal{C}_{\beta }\mathcal{C}_{\gamma }-\frac{1}{2}%
\delta _{\alpha \beta }\mathcal{C}_{\gamma }-\frac{1}{2}%
\delta _{\alpha \gamma }\mathcal{C}_{\beta }-\frac{1}{2}%
\delta _{\beta \gamma }\mathcal{C}_{\alpha })$ and so on. The dimensionless
particle velocity is $\vec{\mathcal{C}}=\left( \frac{m}{2kT}\right) ^{1/2}%
\vec{c}$. These tensors are orthogonal in the Hilbert space $\mathcal{H},$
satisfying 
\begin{equation}
\int e^{-\mathcal{C}^{2}}\Psi _{\theta ,(r_{_{\theta }})}\Psi _{\theta
,(s_{_{\theta }})}d\vec{\mathcal{C}}=\lambda _{\theta }\Delta _{(r_{_{\theta
}})(s_{_{\theta }})},
\end{equation}%
where $\Delta _{(r_{_{\theta }})(s_{_{\theta }})}^{2\theta }$ is a $2\theta $%
-order isotropic tensor, \cite{wolfram}, and $\lambda _{\theta }$ is a
constant. With $\vec{\mathcal{U}}=\left( \frac{m}{2kT}\right) ^{1/2}\vec{u}$%
, the coefficients $a_{\theta ,(r_{_{\theta }})}^{eq}$ in Eq. (\ref{serieseq}%
) are the moments $a_{_{0}}^{eq}=n$, $a_{_{1,\alpha }}^{eq}=n\mathcal{U}%
_{\alpha }$, $a_{_{2,\alpha \beta }}^{eq}=n\mathcal{U}_{\alpha }\mathcal{U}%
_{\beta }$, $a_{_{3,\alpha \beta \gamma }}^{eq}=n\mathcal{U}_{\alpha }%
\mathcal{U}_{\beta }\mathcal{U}_{\gamma }$ and so on, which are dependents
on the volumetric number of particles $n$, on the dimensionless macroscopic
velocity $\vec{\mathcal{U}}$ and on the temperature $T$.\newline
For each point $\vec{x}$ the distribution function $\phi $ in the
non-equilibrium part $f^{neq}$ $=f^{eq}\phi $ can be developed in terms of
the orthogonal basis $\Psi _{\theta ,(r_{_{\theta }})}$, \cite{philippi}, 
\cite{shanhe}, written in terms of the velocity fluctuation $\vec{\mathcal{C}%
}_{f}=\frac{\vec{c}-\vec{u}}{\left( \frac{2kT}{m}\right) ^{1/2}}=\frac{\vec{C%
}}{\left( \frac{2kT}{m}\right) ^{1/2}}=\vec{\mathcal{C}}-\vec{\mathcal{U}}$ 
\begin{equation}
\phi =\sum_{\theta }a_{\theta ,(r_{_{\theta }})}^{\phi }\left( \vec{x}%
,t\right) \Psi _{\theta ,(r_{_{\theta }})}\left( \vec{\mathcal{C}}%
_{f}\right) ,  \label{fneq}
\end{equation}%
and coefficients $a_{\theta }^{\phi }$ can be related to the macroscopic
moments of $f$. In this way, $a_{0}^{\phi }=0,$ $a_{1,\alpha }^{\phi }=0$.
The coefficient $a_{2,\alpha \beta }^{\phi }$ is related to the viscous
stress tensor ${\LARGE \tau }_{\alpha \beta }$ through 
\begin{equation}
a_{2,\alpha \beta }^{\phi }=\frac{{\LARGE \tau }_{\alpha \beta }}{2P},
\label{tau2}
\end{equation}%
where $P=nkT$ is the thermodynamic pressure.\newline
The fluctuation kinetic energy $E(\vec{x},t)$ is given by 
\begin{equation}
E(\vec{x},t)=\int f\frac{1}{2}m\left( \vec{c}- \vec{u}\right) ^{2}d\vec{c}%
=\int f^{eq}\frac{1}{2}m\left( \vec{c}-\vec{u}\right) ^{2}d\vec{c.}
\end{equation}
In this way 
\begin{equation}
\int f^{neq}\frac{1}{2}m\left( C\right) ^{2}d\vec{C}=0,
\end{equation}%
or 
\begin{equation}
\int f^{neq}\frac{1}{2}mC_{\alpha }C_{\alpha }d\vec{C}=\frac{1}{2}tr\left( 
{\LARGE \tau }\right) =0.
\end{equation}
In two-dimensions 
\begin{equation}
{\LARGE \tau }_{xx}+{\LARGE \tau }_{yy}=0,  \label{tau0}
\end{equation}%
or 
\begin{equation}
a_{2,xx}^{\phi }+a_{2,yy}^{\phi }=0.
\end{equation}
For third-order moments 
\begin{eqnarray}
S_{\alpha \beta \gamma } &=&\int fmc_{\alpha }c_{_{\beta }}c_{\gamma }d\vec{c%
}=\int f^{eq}mc_{\alpha }c_{_{\beta }}c_{\gamma }d\vec{c}+\int
f^{neq}mc_{\alpha }c_{_{\beta }}c_{\gamma }d\vec{c}  \notag \\
&=&S_{\alpha \beta \gamma }^{eq}+S_{\alpha \beta \gamma }^{neq},
\end{eqnarray}
with 
\begin{equation}
S_{\alpha \beta \gamma }^{eq}=\rho u_{\alpha }u_{\beta }u_{\gamma }+P\left(
\delta _{\beta \gamma }u_{\alpha }+\delta _{\alpha \gamma }u_{\beta }+\delta
_{\alpha \beta }u_{\gamma }\right) .  \label{seq}
\end{equation}
For the non-equilibrium part, 
\begin{equation}
S_{\alpha \beta \gamma }^{neq}=\int f^{neq}mc_{f\alpha}c_{f\beta }c_{f\gamma }d%
\vec{c}+\left( {\LARGE \tau }_{\beta \gamma }u_{\alpha }+{\LARGE \tau }%
_{\alpha \gamma }u_{\beta }+{\LARGE \tau }_{\alpha \beta }u_{\gamma }\right)
,
\end{equation}%
resulting, using $a_{1,\alpha }^{\phi }=0$, the invariance property with
respect to index permutation and Eq. (\ref{seq}): 
\begin{eqnarray}
P\left( \frac{2kT}{m}\right) ^{\frac{1}{2}}a_{3,\alpha \beta \gamma }^{\phi
} &=&\frac{S_{\alpha \beta \gamma }}{2}-\left[ 
\begin{array}{c}
\frac{1}{2}\rho u_{\alpha }u_{\beta }u_{\gamma }+\frac{1}{2}P\left( \delta
_{\beta \gamma }u_{\alpha }+\delta _{\alpha \gamma }u_{\beta }+\delta
_{\alpha \beta }u_{\gamma }\right) \\ 
+\frac{1}{2}\left( {\LARGE \tau }_{\beta \gamma }u_{\alpha }+{\LARGE \tau }%
_{\alpha \gamma }u_{\beta }+{\LARGE \tau }_{\alpha \beta }u_{\gamma }\right)%
\end{array}%
\right]  \notag \\
&\equiv &q_{\alpha \beta \gamma }.
\end{eqnarray}
When $\beta $ and $\gamma $ are contracted, defining $\epsilon _{\alpha }$
to be the total energy flux along the direction $\alpha$, 
\begin{equation}
P\left( \frac{2kT}{m}\right) ^{\frac{1}{2}}a_{3,\alpha \beta \beta }^{\phi
}=\epsilon _{\alpha }-\left[ \frac{1}{2}\rho u^{2}u_{\alpha }+P\left( \frac{D%
}{2}+1\right) u_{\alpha }+{\LARGE \tau }_{\alpha \beta }u_{\beta }\right]
=q_{\alpha },
\end{equation}%
where $q_{\alpha }$ is the \textit{net }heat flux along the direction $%
\alpha $, i.e., the total energy flux $\epsilon _{\alpha }$, subtracting
from it, the flow of macroscopic kinetic energy $\frac{1}{2}\rho
u^{2}u_{\alpha }$, the compression work $P\left( \frac{D}{2}+1\right)
u_{\alpha }$ and the viscous work ${\LARGE \tau }_{\alpha \beta }u_{\beta }$.

\subsection{Collision term}

Particles are supposed to be material points without volume and only able to
exchange translational kinetic energy, but the collision term $\Omega $ is,
here, considered to take multiparticles collisions into account. Since, in
this case, the collision term structure is not known, some assumptions are
required. In this manner, near the equilibrium, $\Omega $ is considered to
be $f^{eq}\mathcal{L(}\phi \mathcal{)}$, the operator $\mathcal{L}$ being a
linear operator. This property was shown to be true for binary collisions, $%
\left[ 14\right] $ and is, here, extended for multiparticles collisions.
When $f$ \ is near $f^{eq}$, Boltzmann equation reads 
\begin{equation}
\partial _{t}f+\vec{c}.\nabla f=\Omega =f^{eq}\mathcal{L(}\phi ).
\end{equation}
Using the development, Eq. $\left( \ref{fneq}\right) ,$ 
\begin{equation}
\mathcal{L(}\phi )=\sum_{\theta }a_{\theta ,(r_{\theta })}^{\phi }\mathcal{L}%
\left( \Psi _{\theta ,(r_{\theta })}\right) .
\end{equation}
The $\theta $-order tensor $\mathcal{L}\left( \Psi _{\theta ,(r_{\theta
})}\right) $ is, itself, an element of the $\mathcal{C}^{D}$ space and can
be developed in terms of \ the $\theta $-order Hermite tensors that belong
to the orthogonal basis of this \ space, 
\begin{equation}
\mathcal{L}\left( \Psi _{_{\theta ,(r_{\theta })}}\right) =\sum_{(s_{\theta
})}\gamma _{_{(r_{\theta }),(s_{\theta })}}\Psi _{_{\theta ,(s_{\theta })}},
\end{equation}%
where $\gamma _{(r_{\theta }),(s_{\theta })}$ designate the $(r_{\theta
}),(s_{\theta })$ components of $2\theta $-order \textit{relaxation tensors}%
. Considering, as for binary collision, $\mathcal{L}$ to be a self-adjoint
operator, with non-positive eigenvalues, 
\begin{equation}
\gamma _{_{(r_{\theta }),(m_{\theta })}}=\frac{\int e^{-\mathcal{C}_{f}^{2}}%
\mathcal{L}\left( \Psi _{_{\theta ,(r_{\theta })}}\right) \Psi _{_{\theta
,(m_{\theta })}}d\vec{\mathcal{C}}_{f}}{\int e^{-\mathcal{C}_{f}^{2}}\left(
\Psi _{_{\theta ,(m_{\theta })}}\right) ^{2}d\vec{\mathcal{C}}_{f}}\leq 0.
\end{equation}
Using Einstein\'s notation 
\begin{equation}
\mathcal{L(}\phi )=\sum_{\theta }\gamma _{_{(r_{\theta }),(s_{\theta
})}}a_{\theta ,(r_{\theta })}^{\phi }\Psi _{_{\theta ,(s_{\theta })}},
\label{model series}
\end{equation}
where repeated indexes mean summation.\newline
Above equation is an infinite summation on $\theta $. When the terms above a
chosen order N are diagonalised, following a Gross-Jackson procedure, \cite%
{cercignani}, 
\begin{equation}
\mathcal{L(}\phi )=\sum_{\theta =0}^{N}\gamma _{_{(r_{\theta }),(s_{\theta
})}}a_{\theta ,(r_{\theta })}^{\phi }\Psi _{_{\theta ,(s_{\theta })}}-\gamma
_{_{N+1}}\sum_{\theta =N+1}^{\infty }\delta _{_{(r_{\theta }),(s_{\theta
})}}a_{\theta ,(r_{\theta })}^{\phi }\Psi _{_{\theta ,(s_{\theta })}},
\label{lfneq1}
\end{equation}%
where 
\begin{equation}
\delta _{_{(r_{\theta }),(s_{\theta })}}=\delta
_{r_{_{1}}s_{_{1}}}....\delta _{r_{_{\theta }}s_{_{\theta }}}.
\end{equation}
In this way, using Eq. $\left( \ref{fneq}\right) $ 
\begin{equation}
\mathcal{L(}\phi )=-\left[ \sum_{\theta =0}^{N}\lambda _{_{(r_{_{\theta
}}),(s_{_{\theta }})}}a_{\theta ,(r_{_{\theta }})}^{\phi }\Psi _{_{\theta
,(s_{_{\theta }})}}\right] -\gamma _{_{N+1}}\phi ,  \label{model}
\end{equation}%
where $\lambda _{_{(r_{_{\theta }}),(s_{_{\theta }})}}=-\left( \gamma
_{_{(r_{\theta }),(s_{\theta })}}+\gamma _{_{N+1}}\delta _{_{(r_{\theta
}),(s_{\theta })}}\right) $ is positive for all $r_{\theta },s_{\theta }$,
since a) $\lambda _{_{(r_{_{\theta }}),(s_{_{\theta }})}}=-\gamma
_{_{(r_{\theta }),(s_{\theta })}}$ for all off-diagonal components and b)
the diagonal components $\gamma _{_{(r_{\theta }),(r_{\theta })}}$ are
negative with an absolute value that is greater than $\gamma _{_{N+1}}$ for
all $\theta $ smaller or equal to $N$. Eq. (\ref{model}) can be considered
as an N$^{th}$-order kinetic model to the collision term, with an absorption
term $\gamma _{_{N}}\phi $ resulting from the diagonalization of the
relaxation tensors after the given $N$. Therefore, all the moments of order
higher than $N$ are collapsed into a single non-equilibrium term minimizing
the truncation effects on the fine structure of the operator $\mathcal{L}$
spectrum.\newline
Although very little is known about the \textit{true} collision term $\Omega 
$ when multiple collisions are considered, Eq. (\ref{model}) generates
increasing accuracy models to $\Omega $ when the distribution function $f$ \
is near the Maxwell-Boltzmann equilibrium distribution, $\ f^{eq}$. The only
restrictions are: a) particles were considered as material points without
volume and b) particles internal energy and long-range forces among the
particles were not considered in the derivation.\newline
When $N=0$ or $N=1$, Eq. (\ref{model}) gives the well known BGK model, when
all the collision operator spectra is replaced by a single relaxation term.%
\newline
Each term in the sum, in Eq. (\ref{model}), gives the relaxation to the
equilibrium of second or higher order kinetic moments M$_{\theta }$ that are
not preserved in collisions, modulated by a $\lambda _{\theta }$ relaxation
tensor. In Section 1.3, explicit expressions are given for the collision
models. When the diagonalization constant is considered to be zero, i.e.,
when the series, Eq. (\ref{model series}), is truncated above N, replacing $%
\Omega =f^{eq}\mathcal{L(}\phi )$ in the Boltzmann equation, the inner
products of the resulting equation by $\Psi _{_{\chi ,(s_{\chi })}}$ give 
\begin{equation}
\partial _{t}a_{\chi ,(r_{\chi })}^{f}+\vec{c}.\nabla a_{\chi ,(r_{\chi
})}^{f}=\lambda _{(r_{\chi })(s_{\chi })}a_{\chi ,(s_{\chi })}^{neq},
\label{dhumieres}
\end{equation}%
where the distribution function \ $f$\ was developed following 
\begin{equation}
f=\sum_{\theta }e^{-\mathcal{C}_{f}^{2}}a_{\theta ,(r_{\theta })}^{f}\Psi
_{\theta ,(r_{\theta })},
\end{equation}%
and 
\begin{equation}
a_{\theta ,(r_{\theta })}^{neq}=n\left( \frac{m}{2\pi kT}\right)
^{D/2}a_{\theta ,(r_{\theta })}^{\phi }.
\end{equation}
It can be easily seen that D\'Humi\`{e}res moment equations (\cite{dhumieres}%
, \cite{dhumieresetal}) are particular discrete forms of Eq. (\ref{dhumieres}%
). Nevertheless, in d\'Humi\`{e}res moments method, \textit{all }the
b-moments in a b-discrete velocities set are considered. It was shown in 
\cite{heluo}, that the number of degrees of freedom of a given lattice
restricts the order $n$ of the kinetic moments with exact quadrature. This
means that all the moments which order are greater than $n$ cannot be
correctly described in this given lattice. As it was mentioned in the
Introduction, in the moments method these \textit{high-frequency }moments
are forced to give consistent and numerical stable \textit{low-frequency}
macroscopic equations by using dispersion relations, decreasing the effect
of numerical instability sources, but buffering the appearance of complex
flow structures, when the Reynolds number increases.

\subsection{Collision models for the continuous Boltzmann equation}

In present section, the isotropy of 4$^{th}$ and 6$^{th}$ rank tensors will
be used to give explicit forms for the second and third-order collision
models, Eq. (\ref{model}). Without any loss in the generality, we restrict
ourselves to two-dimensional spaces.

\subsubsection{Second order model in the two-dimensional space}

From Eq. (\ref{model})%
\begin{equation}
\lambda _{_{(r_{_{2}}),(s_{_{2}})}}a_{2,(r_{2})}^{\phi }\Psi
_{_{2,(s_{2})}}=\lambda _{_{\alpha \beta \gamma \delta }}a_{2,\alpha \beta
}^{\phi }\Psi _{_{2,\gamma \delta }}.
\end{equation}

Requiring isotropy of 4$^{th}$ rank tensors and considering the symmetry
with respect to index permutation,

\begin{equation}
\lambda _{_{\alpha \beta \gamma \delta }}=\lambda _{\mu }\left( \delta
_{\alpha \beta }\delta _{\gamma \delta }+\delta _{\alpha \gamma }\delta
_{\beta \delta }+\delta _{\alpha \delta }\delta _{\beta \gamma }\right) .
\end{equation}

In this way,

\begin{eqnarray}
\lambda _{_{\alpha \beta \gamma \delta }}a_{2,\alpha \beta }^{\phi }\Psi
_{_{2,\gamma \delta }} &=&\lambda _{\mu }\left[ a_{2,\alpha \alpha }^{\phi
}\Psi _{_{2,\gamma \gamma }}+a_{2,\alpha \beta }^{\phi }\Psi _{_{2,\alpha
\beta }}+a_{2,\alpha \beta }^{\phi }\Psi _{_{2,\beta \alpha }}\right]  \notag
\\
&=&\lambda _{\mu }\left[ 
\begin{array}{c}
a_{2,xx}^{\phi }\left( \mathcal{C}_{fx}^{2}-\frac{1}{2}\right)
+a_{2,yy}^{\phi }\left( \mathcal{C}_{fy}^{2}-\frac{1}{2}\right) + \\ 
2a_{2,xy}^{\phi }\mathcal{C}_{fx}\mathcal{C}_{fy}%
\end{array}%
\right] ,
\end{eqnarray}%
since $a_{2,\alpha \alpha }^{\phi }=0$. Using Eq.(\ref{tau2})

\begin{equation}
\lambda _{_{\alpha \beta \gamma \delta }}a_{2,\alpha \beta }^{\phi }\Psi
_{_{2,\gamma \delta }}=\frac{\lambda _{\mu }}{P}\left[ {\LARGE \tau }%
_{xx}\left( \mathcal{C}_{fx}^{2}-\frac{1}{2}\right) +{\LARGE \tau }%
_{yy}\left( \mathcal{C}_{fy}^{2}-\frac{1}{2}\right) +2{\LARGE \tau }_{xy}%
\mathcal{C}_{fx}\mathcal{C}_{fy}\right] ,
\end{equation}%
or, from Eq. (\ref{tau0}), ${\LARGE \tau }_{xx}=-{\LARGE \tau }_{yy}$

\begin{equation}
\lambda _{_{\alpha \beta \gamma \delta }}a_{2,\alpha \beta }^{\phi }\Psi
_{_{2,\gamma \delta }}=\frac{\lambda _{\mu }}{P}\left[ {\LARGE \tau }%
_{xx}\left( \mathcal{C}_{fx}^{2}-\mathcal{C}_{fy}^{2}\right) +2{\LARGE \tau }%
_{xy}\mathcal{C}_{fx}\mathcal{C}_{fy}\right] .
\end{equation}

Second order model in two dimensions will be written as%
\begin{eqnarray}
\mathcal{L}^{(2)}\mathcal{(}\phi ) &=&-\frac{\lambda _{\mu }}{P}\left[ 
{\LARGE \tau }_{xx}\left( \mathcal{C}_{fx}^{2}-\frac{1}{2}\right) +{\LARGE %
\tau }_{yy}\left( \mathcal{C}_{fy}^{2}-\frac{1}{2}\right) +2{\LARGE \tau }%
_{xy}\mathcal{C}_{fx}\mathcal{C}_{fy}\right] -  \notag \\
&&\gamma _{_{3}}\phi .  \label{2ndorder}
\end{eqnarray}

\subsubsection{\protect\bigskip Third-order model}

From Eq. (\ref{model})%
\begin{equation}
\lambda _{_{(r_{_{3}}),(s_{_{3}})}}a_{3,(r_{3})}^{\phi }\Psi
_{_{3,(s_{3})}}=\lambda _{_{\alpha \beta \gamma \delta \zeta \eta
}}a_{3,\alpha \beta \gamma }^{\phi }\Psi _{_{3,\delta \zeta \eta }}.
\end{equation}

For isotropic fluids, tensor $\lambda _{_{\alpha \beta \gamma \delta \zeta
\eta }}$ is a linear combination of five 6$^{th}$ order tensors given by the
recurrence relation, \cite{wolfram},

\begin{equation}
\Delta _{r_{1}....r_{6}}^{(6)}=\delta _{r_{1}r_{j}}\Delta
_{r_{1}...r_{j-1},r_{j+1}....r_{4}}^{(4)},
\end{equation}%
resulting

\begin{equation}
\lambda _{_{\alpha \beta \gamma \delta \zeta \eta }}a_{3,\alpha \beta
\gamma }^{\phi }\Psi _{_{3,\delta \zeta \eta }}  
=\lambda _{_{1}}a_{3,\alpha \beta
\gamma }^{\phi }\Psi _{_{3,\alpha \beta \gamma }}+\lambda _{_{2}}a_{3,\alpha \beta
\beta }^{\phi }\Psi _{_{3,\alpha \beta \beta }}
\end{equation}%

\subsection{Macroscopic thermohydrodynamic equations}

Macroscopic thermohydrodynamic equations may be obtained from the Boltzmann
equation by multiplying this equation by the mass m, the momentum m$\vec{c}$
and the kinetic energy$\ \frac{1}{2}mc^{2}$ of the particles and integrating
the resulting equations in the $\vec{c}$ velocity space.\newline
The mass conservation reads, as usually,

\begin{equation}
\partial _{t}\rho +\partial _{\alpha }\left( \rho u_{\alpha }\right) =0.
\label{masscons}
\end{equation}

>From the momentum preservation in collisions

\begin{equation}
\partial _{t}\left( \rho u_{\alpha }\right) +\partial _{\alpha }\left( \rho
u_{\alpha }u_{\beta }+P\delta _{\alpha \beta }+{\LARGE \tau }_{\alpha \beta
}\right) =0,  \label{mompres}
\end{equation}%
where $P$ is the thermodynamics pressure, $P=nkT$ and ${\LARGE \tau }%
_{\alpha \beta }$ is the viscous stress tensor.\newline
When Eq.$\left( \text{\ref{mompres}}\right) $ is multiplied by $u_{\alpha }$%
, the macroscopic kinetic energy, $\frac{1}{2}\rho u^{2}$, balance equation
is obtained

\begin{equation}
\partial _{t}\left( \text{$\frac{1}{2}$}\rho u_{\alpha }^{2}\right)
=P\triangledown .u+{\LARGE \tau }_{\alpha \beta }\partial _{\beta }u_{\alpha
}-\partial _{\beta }\left( \text{$\frac{1}{2}$}\rho u_{\alpha }^{2}u_{\beta
}+P\delta _{\alpha \beta }u_{\alpha }+{\LARGE \tau }_{\alpha \beta
}u_{\alpha }\right) .  \label{kinenerbal}
\end{equation}%
\ 

The total energy conservation equation reads

\begin{equation}
\partial _{t}\left( E+\text{$\frac{1}{2}$}\rho u_{\alpha }^{2}\right)
=-\partial _{\beta }\left[ \left( \text{$\frac{1}{2}$}\rho u_{\alpha
}^{2}+E\right) u_{\beta }+\left( P\delta _{\alpha \beta }+{\LARGE \tau }%
_{\alpha \beta }\right) u_{\alpha }+q_{\beta }\right] ,  \label{totenerg}
\end{equation}%
where E is the thermodynamics internal energy

\begin{equation}
E=\int \frac{1}{2}m\left( \vec{c}-\vec{u}\right) ^{2}fd\vec{c}=\int \frac{1}{%
2}m\left( \vec{c}-\vec{u}\right) ^{2}f^{eq}d\vec{c}=\frac{D}{2}nkT.
\end{equation}

The internal energy balance equation is obtained by subtracting Eq. (\ref%
{kinenerbal}) from Eq. (\ref{totenerg}),

\begin{equation}
\partial _{t}\left( E\right) =-\left( P\triangledown .u+{\LARGE \tau }%
_{\alpha \beta }\partial _{\beta }u_{\alpha }\right) -\partial _{\beta }%
\left[ Eu_{\beta }+q_{\beta }\right] ,  \label{intenrgbal}
\end{equation}%
where $-\left( P\triangledown .u+{\LARGE \tau }_{\alpha \beta }\partial
_{\beta }u_{\alpha }\right) $ is the source term of internal energy.

Equations (\ref{masscons}, \ref{mompres} and \ref{intenrgbal}) form a closed
set of equations when \ the viscous stress tensor ${\LARGE \tau }_{\alpha
\beta }$ and the heat flux vector $q_{\beta }$ are known in terms of the
spatial gradients of the first macroscopic moments, $\rho $, $\vec{u}$ and $%
T $ of the distribution function. This is accomplished when the Knudsen
number, $\ Kn\longrightarrow 0$, \ by performing a Chapman-Enskog asymptotic
analysis of the modelled Boltzmann equation.

\subsection{Chapman Enskog analysis for the continuous model}

Considering $f^{0}$ in the asymptotic expansion

\begin{equation}
f=f^{0}+Knf^{1}+...,
\end{equation}%
to be the Maxwell-Boltzmann equilibrium distribution $f^{eq}(n,\vec{u},T)$,
the zeroth order time derivative resulting from the Chapman-Enskog induced
decomposition of the time derivative reads,

\begin{eqnarray}
\frac{1}{f^{0}}\frac{d_{0}f^{0}}{dt} &=&2\left( \mathcal{C}_{f\alpha }%
\mathcal{C}_{f\beta }-\frac{1}{2}\delta _{\alpha \beta }\right) \partial
_{\beta }u_{\alpha }-\frac{2}{D}\left( \mathcal{C}_{f\alpha }^{2}-\frac{D}{2}%
\right) \triangledown .\vec{u}  \notag \\
&&+\left( \frac{2kT}{m}\right) ^{1/2}\left( \mathcal{C}_{f}^{2}-\frac{D+2}{2}%
\right) \vec{\mathcal{C}}_{f}.\triangledown \ln T.  \label{df0}
\end{eqnarray}

\subsubsection{Second order model in two dimensions}

Using Eqs. (\ref{df0} and \ref{2ndorder})

\begin{eqnarray}
&&2\left( \mathcal{C}_{fx}^{2}-\frac{1}{2}\right) \partial _{x}u_{x}+2\left( 
\mathcal{C}_{fy}^{2}-\frac{1}{2}\right) \partial _{y}u_{y}+2\mathcal{C}_{fx}%
\mathcal{C}_{fy}\left( \partial _{x}u_{y}+\partial _{y}u_{x}\right)  \notag
\\
&&-\left[ \left( \mathcal{C}_{fx}^{2}-\frac{1}{2}\right) +\left( \mathcal{C}%
_{fy}^{2}-\frac{1}{2}\right) \right] \triangledown .\vec{u}+\left( \frac{2kT%
}{m}\right) ^{1/2}\left( \mathcal{C}_{f}^{2}-2\right) \vec{\mathcal{C}}%
_{f}.\triangledown \ln T  \notag \\
&=&-\frac{2\lambda _{\mu }}{P}\left[ {\LARGE \tau }_{xx}\left( \mathcal{C}%
_{fx}^{2}-\frac{1}{2}\right) +{\LARGE \tau }_{yy}\left( \mathcal{C}_{fy}^{2}-%
\frac{1}{2}\right) +2{\LARGE \tau }_{xy}\mathcal{C}_{fx}\mathcal{C}_{fy}%
\right] -\gamma _{_{3}}\phi .  \label{2nd2d}
\end{eqnarray}

For finding the correct expression of ${\LARGE \tau }_{\alpha \beta }$, in
terms of the spatial derivatives of the macroscopic variables, the inner
product of the above equation by $\left( \mathcal{C}_{f\alpha }\mathcal{C}%
_{f\beta }-\frac{1}{2}\delta _{\alpha \beta }\right) $ in the $\mathcal{C}_{f%
\text{ }}$ velocity space is performed. By multiplying the above equation by 
$\left( \mathcal{C}_{fx}^{2}-\frac{1}{2}\right) $

\begin{equation}
\left( \lambda _{\mu }+\gamma _{_{3}}\right) \frac{{\LARGE \tau }_{xx}}{P}%
=-\partial _{x}u_{x}+\frac{1}{2}\triangledown .\vec{u}=\frac{1}{2}\left(
\partial _{y}u_{y}-\partial _{x}u_{x}\right) .
\end{equation}

Similarly

\begin{equation}
\left( \lambda _{\mu }+\gamma _{_{3}}\right) \frac{{\LARGE \tau }_{yy}}{P}=%
\frac{1}{2}\left( \partial _{x}u_{x}-\partial _{y}u_{y}\right) ,
\end{equation}%
and

\begin{equation}
\left( \lambda _{\mu }+\gamma _{_{3}}\right) \frac{{\LARGE \tau }_{xy}}{P}=-%
\frac{1}{2}\left( \partial _{x}u_{y}+\partial _{y}u_{x}\right) .
\end{equation}

These results give for the first and second viscosity coefficients$,$

\begin{equation}
\mu =\eta =\frac{nkT}{2\lambda _{\mu }+\gamma _{_{3}}},
\label{visc}
\end{equation}%
in the relation 
\begin{equation}
{\LARGE \tau }_{\alpha \beta }=-\mu \left( \partial _{\alpha }u_{\beta
}+\partial _{\beta }u_{\alpha }\right) +\eta \triangledown .\vec{u}.
\end{equation}

Eq. (\ref{visc}) means that the first and the second viscosity coefficients
are not independent quantities and this result must be considered as a
limitation resulting from the continuous collision term itself, \ where the
particles were considered as material points with translational degrees of
freedom. In fact, this same result will be retrieved when using the third or
higher order model for the collision term. The consideration of internal
energy modes would be necessary for an up-grade of Eq. (\ref{visc}), \cite%
{philippi}.\newline
The third-order moment $a_{3,xxx}^{\phi }$ may be obtained by multiplying
Eq. (\ref{2nd2d}) by $\left( \mathcal{C}_{fx}^{2}-\frac{3}{2}\right) 
\mathcal{C}_{fx}$. In this way,

\begin{equation}
a_{3,xxx}^{\phi }=-\frac{3}{4}\frac{\left( \frac{2kT}{m}\right) ^{1/2}}{%
\gamma _{_{3}}}\partial _{x}\ln T.
\end{equation}

Multiplying Eq. (\ref{2nd2d}) by $\left( \mathcal{C}_{fy}^{2}-\frac{1}{2}%
\right) \mathcal{C}_{fx}$,

\begin{equation}
a_{3,yyx}^{\phi }=-\frac{1}{4}\frac{\left( \frac{2kT}{m}\right) ^{1/2}}{%
\gamma _{_{3}}}\partial _{x}\ln T.
\end{equation}

But

\begin{equation}
q_{x}=P\left( \frac{2kT}{m}\right) ^{1/2}\left( a_{3,xxx}^{\phi
}+a_{3,yyx}^{\phi }\right) =-\frac{P\left( \frac{2kT}{m}\right) }{\gamma
_{_{3}}T}\partial _{x}T,
\end{equation}%
giving for the thermal conductivity

\begin{equation}
K=\frac{\left(D+2\right)nk^{2}T}{2m}\frac{1}{\gamma _{_{3}}}.
\end{equation}

In this manner, present second-order continuous kinetic model is
thermodynamic consistent and able for analyzing non-isothermal and fully
compressible flows. The thermal conductivity is related to $\gamma _{_{3}}$
diagonalization constant. Consideration of third-order models will be, only,
necessary in multi-component systems, for correctly describing third-order
coupling: the Soret and Dufour effects, \cite{philippi}.

\section{Discretization}

In present sections an analysis is performed, trying to emphasize the 
\textit{theoretically identifiable} effects of time and velocity
discretization on the ability of the derived discrete models in retrieving
the correct thermohydrodynamic equations, i.e., the full compressible
Navier-Stokes equations and the thermodynamic internal energy balance
equation, with the Fourier heat flux term.

\subsection{Time discretization}

Boltzmann equation with the kinetic model Eq. (\ref{model})becomes

\begin{equation}
\frac{d}{dt}f+\gamma _{_{N}}f=f^{eq}\sum_{\theta =0}^{N}\lambda
_{_{(r_{_{\theta }}),(s_{_{\theta }})}}a_{\theta ,(r_{_{\theta }})}^{\phi
}\Psi _{_{\theta ,(s_{_{\theta }})}}+\gamma _{_{N}}f^{eq}.
\end{equation}

For avoiding time-step errors $\mathcal{O}\left( \delta ^{2}\right) $,
Boltzmann equation is integrated between $t$\ and $t+\delta $, considering
linear approximations for $f^{eq}\left( \vec{x}+\vec{c}t^{\prime },\vec{c}%
,t+t^{\prime }\right) $ and, also, for $a_{\theta ,(r_{_{\theta }})}^{neq}$,
since $a_{\theta ,(r_{_{\theta }})}^{neq}\left( \vec{x}+\vec{c}t^{\prime
},t+t^{\prime }\right) $, when $0\leq t^{\prime }\leq \delta $, \cite%
{hechendoolen}. The result is

\begin{eqnarray}
&&f\left( \vec{x}+\vec{c}\delta ,\vec{c},t+\delta \right) -f\left( \vec{x},%
\vec{c},t\right)  \notag \\
&=&\left( \gamma _{N}\delta \right) \frac{1}{2}\left[ f^{eq}\left( \vec{x}+%
\vec{c}\delta ,\vec{c},t+\delta \right) +f^{eq}\left( \vec{x},\vec{c}%
,t\right) \right]  \notag \\
&&-\left( \gamma _{N}\delta \right) \frac{1}{2}\left[ f\left( \vec{x}+\vec{c}%
\delta ,\vec{c},t+\delta \right) +f\left( \vec{x},\vec{c},t\right) \right] 
\notag \\
&&+\sum_{\theta =0}^{N}\left( \delta \lambda _{_{(r_{_{\theta
}}),(s_{_{\theta }})}}\right) \Psi _{_{\theta ,(s_{_{\theta
}})}}f^{eq}\left( \vec{x},\vec{c},t\right) \text{ X}  \notag \\
&&\text{X }\frac{1}{2}\left[ a_{\theta ,(r_{_{\theta }})}^{\phi }\left( \vec{%
x}+\vec{c}\delta ,t+\delta \right) +a_{\theta ,(r_{_{\theta }})}^{\phi
}\left( \vec{x},t\right) \right] .
\end{eqnarray}

This corresponds to an implicit numerical scheme (in fact, a Crank-Nicholson
scheme). Although implicit schemes are very easily manageable in the
lattice-Boltzmann context, \cite{krafk}, \cite{sanka}, if one wants to avoid
implicitness a new distribution can be defined as,

\begin{equation}
\widetilde{f}=f+\gamma _{_{N}}\frac{1}{2}\delta \left( f-f^{eq}\right)
-\sum_{\theta =0}^{N}\frac{\delta }{2}\lambda _{_{(r_{_{\theta
}}),(s_{_{\theta }})}}a_{\theta ,(r_{_{\theta }})}^{\phi }\Psi _{_{\theta
,(s_{_{\theta }})}}f^{eq},  \label{ftil}
\end{equation}%
resulting

\begin{eqnarray}
&&\widetilde{f}\left( \vec{x}+\vec{c}\delta ,\vec{c},t+\delta \right)  \notag
\\
=\widetilde{f}\left( \vec{x},\vec{c},t\right) + &&\frac{\delta }{\tau
_{_{N}}+\frac{1}{2}\delta }\left( f^{eq}-\widetilde{f}\right) +  \notag \\
&&\frac{\delta }{\tau _{_{N}}+\frac{1}{2}\delta }\sum_{\theta =0}^{N}\tau
_{_{N}}\lambda _{_{(r_{_{\theta }}),(s_{_{\theta }})}}a_{\theta
,(r_{_{\theta }})}^{\phi }f^{eq}\Psi _{_{\theta ,(s_{_{\theta }})}},
\label{model1}
\end{eqnarray}%
where $\tau _{_{N}}=\frac{1}{\gamma _{N}}$. It must be observed that $%
a_{\theta ,(r_{_{\theta }})}^{\phi }$ are the macroscopic moments of $f$ and
not of $\widetilde{f}$. Nevertheless, it can be shown from Eq. (\ref{ftil})
that $\widetilde{a}_{\theta ,(s_{_{\theta }})}^{\phi }$ and $a_{\theta
,(r_{_{\theta }})}^{\phi }$ are directly related by

\begin{equation}
\widetilde{a}_{\theta ,(s_{_{\theta }})}^{\phi }=\left( 1+\frac{\delta }{%
2\tau _{_{N}}}\right) a_{\theta ,(s_{_{\theta }})}^{\phi }-\frac{\delta }{2}%
\lambda _{_{(r_{_{\theta }}),(s_{_{\theta }})}}a_{\theta ,(r_{_{\theta
}})}^{\phi },  \label{atil}
\end{equation}

Consider, for instance, the second order model. In this case, it can be
shown that

\begin{eqnarray}
\lambda _{_{(r_{_{\theta }}),(s_{_{\theta }})}}a_{\theta ,(r_{_{\theta
}})}^{\phi } &=&\lambda _{_{\alpha \beta \gamma \delta }}a_{2,\alpha \beta
}^{\phi }=-\lambda _{\mu }\left( \delta _{\alpha \beta }\delta _{\gamma
\delta }+\delta _{\alpha \gamma }\delta _{\beta \delta }+\delta _{\alpha
\delta }\delta _{\beta \gamma }\right) a_{2,\alpha \beta }^{\phi }  \notag \\
&=&-2\lambda _{\mu }\left( a_{2,\gamma \delta }^{neq}\right) ,
\end{eqnarray}%
since $a_{2,\alpha \alpha }^{neq}=0$ and $\lambda _{\mu }$ is required to be
positive.\ Eq. (\ref{atil}) means

\begin{equation}
\widetilde{a}_{2,\gamma \delta }^{\phi }=\left( 1+\frac{\delta }{2\tau
_{_{N}}}+\delta \lambda _{\mu }\right) a_{2,\gamma \delta }^{\phi }.
\end{equation}

\subsection{Velocity discretization}

Velocity discretization is the most critical step in presently proposed
procedure. For athermal problems He and Luo, \cite{heluo}, have shown that
some widely used sets of discrete velocities $\left\{ \vec{c}%
_{i},i=1,...b\right\} $ may be derived from the continuous velocity space by
Gauss-Hermite (D2Q9, D3Q27) and Gauss-Radau (D2Q7) quadrature. All these
sets are space-filling, in the sense that for every lattice site $\vec{x}$, $%
\vec{x}+\vec{c}_{i}$ \ points to another site in the lattice.\newline
If it is agreed that quadrature is the bridge connecting the continuous and
the discrete velocity space, discretization means to replace the entire
continuous velocity space $c^{D}$ by some discrete velocities $\vec{c}_{i}$
satisfying the quadrature for all the kinetic moments of interest, i.e., for
all the kinetic moments that are to be correctly described in
lattice-Boltzmann simulation. Although it is highly desirable set $\vec{c}%
_{i}$ to be space-filling, this condition is not essential for the
discretization itself.\newline
When performing the quadrature, an integration variable must be chosen. If
the dimensionless fluctuation velocity $\vec{\mathcal{C}}_{f}=\frac{\vec{c}-%
\vec{u}}{\left( \frac{2kT}{m}\right) ^{1/2}}$\ \ is chosen as the
integrating variable, considering $\chi $ to be a polynomial of degree $r$
in the velocity,

\begin{equation}
<\chi >=\int f\chi d\vec{c}=n\frac{1}{\pi ^{D/2}}\int e^{-\mathcal{C}%
_{f}^{2}}\chi ^{\prime }\left( \vec{\mathcal{C}}_{f}\right) d\vec{\mathcal{C}%
}_{f}=n\sum_{i=1}^{b}{}\omega _{i}\chi ^{\prime }\left( \vec{\mathcal{C}}%
_{fi}\right) ,
\end{equation}%
where $\vec{\mathcal{C}}_{fi}$ is a discrete velocity (a constant vector),
dependent, basically, on $b$ and on the kind of quadrature operation it is
being performed, $\chi ^{\prime }\left( \vec{\mathcal{C}}_{f}\right) $ is a
polynomial in $\vec{\mathcal{C}}_{f}$ \ of degree $\ r_{p}=r$ $+s$ when it
is related to a preserved moment and $\ r_{p}=r+s+1$, otherwise, \cite{heluo}%
, $s$ being the degree of the polynomial approximation to $f^{eq}$ and $%
\omega _{i}$ are the constant weights to be attributed to each discrete
velocity $\vec{\mathcal{C}}_{fi}$. Exact quadrature restricts the highest
value of \ $r_{p}$ to $r_{p_{m}}$. For a given class of quadrature we can
write $r_{p_{m}}=r_{p_{m}}(b)$, in the sense that increasing $b$ enables
higher degree polynomials to have exact quadrature.\newline
For the first kinetic moment, $n$,

\begin{equation}
n=<1>=n\frac{1}{\pi ^{D/2}}\int e^{-\mathcal{C}_{f}^{2}}1d\vec{\mathcal{C}}%
_{f}=n\sum_{i=1}^{b}\omega _{i}1=n\sum_{i=1}^{b}\omega _{i},
\end{equation}%
resulting,

\begin{equation}
f_{i}^{eq}=\omega _{i}n.
\end{equation}

This means that the discrete equilibrium distribution does not depend, 
\textit{explicitly}, on the macroscopic velocity $\vec{u}$ and on the
temperature $T$. These dependences are included in the particle velocities
through,

\begin{equation}
\vec{c}_{i}=\vec{u}+\left( \frac{2kT}{m}\right) ^{1/2}\vec{\mathcal{C}}_{fi}=%
\vec{c}_{i}\left( T,\vec{u}\right) .
\end{equation}

When the $\vec{\mathcal{C}}_{fi}$ generate a regular lattice, this choice is
possible in LB framework, since LB simulation may, at least in principle, be
performed in the ($\vec{x},\vec{\mathcal{C}}_{fi})$ space. Nevertheless, the 
\textit{physical }grid ($\vec{x},\vec{c}_{i})$, i.e., the physical grid
points where the particles will be located after each time step, will be
time dependent, simulation tends to be very cumbersome and, at a first
sight, boundary conditions will be difficult to satisfy.\newline
Another choice is the dimensionless particle velocity $\vec{\mathcal{C}}=%
\frac{\vec{c}}{\left( \frac{2kT}{m}\right) ^{1/2}}$. This requires to
rewrite the equilibrium distribution as in Eq. (\ref{serieseq}), but, now,
the series%
\begin{equation}
\sum_{\theta }a_{\theta ,(r_{_{\theta }})}^{eq}(n,\vec{\mathcal{U}},T)\Psi
_{\theta ,(r_{_{\theta }})},
\end{equation}%
must be truncated somewhere. This is an important distinguishing point of
discrete models, since in each continuous model presented in Section 1.3,
although the full collision term is replaced by its N$^{th}$-order
approximation, the equilibrium distribution is, always, the full MB
distribution.\newline
Second-order approximations are widely used in athermal simulation, but
thermohydrodynamics require third (or higher) order approximations for the
equilibrium distribution,

\begin{equation}
f^{eq}=n\frac{e^{-\mathcal{C}^{2}}}{\pi ^{D/2}}\left( \frac{m}{2kT}\right)
^{D/2}\left[ 
\begin{array}{c}
1+2\mathcal{C}_{\alpha }\mathcal{U}_{\alpha }+2\left( \mathcal{C}_{\alpha }%
\mathcal{C}_{\beta }-\text{$\frac{1}{2}$}\delta _{\alpha \beta }\right) 
\mathcal{U}_{\alpha }\mathcal{U}_{\beta }+ \\ 
\frac{4}{3}\left( \mathcal{C}_{\alpha }\mathcal{C}_{\beta }-\text{$\frac{3}{2%
}$}\delta _{\alpha \beta }\right) \mathcal{C}_{\gamma }\mathcal{U}_{\alpha }%
\mathcal{U}_{\beta }\mathcal{U}_{\gamma }%
\end{array}%
\right] ,  \label{uexp}
\end{equation}%
which can be viewed as a third-degree polynomial expansion of the $f^{eq}$
dependence on $\vec{\mathcal{U}}$, with errors $O(\mathcal{U}^{4}) $.\newline
After quadrature, the equilibrium distribution becomes

\begin{equation}
f_{i}^{eq}=\omega _{i}n\left( 
\begin{array}{c}
1+2\mathcal{C}_{i\alpha }\mathcal{U}_{\alpha }+2(\mathcal{C}_{i\alpha }%
\mathcal{C}_{i\beta }-\frac{1}{2}\delta _{\alpha \beta })\mathcal{U}_{\alpha
}\mathcal{U}_{\beta } \\ 
+\frac{4}{3}\left( \mathcal{C}_{i\alpha }\mathcal{C}_{i\beta }-\text{$\frac{3%
}{2}$}\delta _{\alpha \beta }\right) \mathcal{C}_{i\gamma }\mathcal{U}%
_{\alpha }\mathcal{U}_{\beta }\mathcal{U}_{\gamma }%
\end{array}%
\right) ,  \label{fiqc1}
\end{equation}%
where, as above, the weights $\omega _{i}$ and the velocity vectors $\vec{%
\mathcal{C}}_{i}$ are dependent on $b$ and on the kind of quadrature that
was performed.\newline
When $\vec{u}=0$, the equilibrium distribution is only dependent on \ the
temperature $T$ \ through the number density of particles, $n$.
Nevertheless, the particle velocities are temperature dependent,

\begin{equation}
\vec{c}_{i}=\left( \frac{2kT}{m}\right) ^{1/2}\vec{\mathcal{C}}_{i}=\vec{c}%
_{i}\left( T\right) .
\end{equation}

A simulation alternative is presented, in this case, by redistributing the
particles among adjacent sites, in accordance with allocation rules, locally
preserving the mass, momentum and kinetic energy of the original packet.
This strategy will be discussed in Section 2.3, leading to the establishment
of temperature dependent velocity models (TDV).\newline
Pavlo \textit{et al., \cite{vahala}, }developed a TDV model based on an
octagonal lattice, which is not space-filling but assures the isotropy of 6$%
^{th}$ rank tensors. It can be shown that this octagonal discrete velocities
set can be retrieved using a Gauss-Radau quadrature, with 8 angular
directions, giving $r_{p_{m}}=7$, instead of 5 as in the D2Q7 model and
assuring the exact quadrature of third-order moments. Additional
considerations can be found in Section 2.4.\newline
Avoiding the $\vec{c}_{i}$ temperature dependence requires to consider the
particles velocity $\vec{c}$ as the integrating variable when performing the
quadrature, i.e., to let $c^{2}$ \ free\ from $T $ in the exponential part $%
e^{-\mathcal{C}^{2}}$of the equilibrium distribution. This can be
accomplished by writing

\begin{equation}
e^{-\frac{\left( c-u\right) ^{2}}{\frac{2kT}{m}}}=\left( e^{-\mathcal{C}%
_{fo}^{2}}\right) ^{\frac{T_{0}}{T}},  \label{dec}
\end{equation}%
where $T_{0}$ is a reference (and constant) temperature and $\ \vec{\mathcal{%
C}}_{fo}=\frac{\vec{c}-\vec{u}}{\left( \frac{2kT_{0}}{m}\right) ^{1/2}}$ is
a new dimensionless fluctuation velocity referred to the temperature $T_{o}$%
. When $T$ \ is near $T_{0}$, i.e., when the departures from thermal
equilibrium are small, the above expression may be developed in a Taylor
series around $\frac{T}{T_{o}}=1$. Considering $\Theta =\frac{T}{T_{o}}-1$,
this development gives

\begin{equation}
\left( e^{-\mathcal{C}_{fo}^{2}}\right) ^{\frac{T_{0}}{T}}=e^{-\mathcal{C}%
_{fo}^{2}}\left[ 1+\mathcal{C}_{fo}^{2}\Theta +\frac{1}{2}\mathcal{C}%
_{fo}^{2}\left( \mathcal{C}_{fo}^{2}-2\right) \Theta ^{2}+...\right] ,
\label{tweights3}
\end{equation}%
which terms are increasing powers of $\mathcal{C}_{fo}^{2}$.\newline
In this way, retaining just the first power in $\theta $,

\begin{eqnarray}
f^{eq} &=&n\left( \frac{T_{0}}{T}\right) ^{D/2}\left[ 1+\mathcal{C}%
_{fo}^{2}\Theta \right] \frac{1}{\pi ^{D/2}}\left( \frac{m}{2kT_{0}}\right)
^{D/2}  \notag \\
&&\times \text{ }e^{-\mathcal{C}_{o}^{2}}\left[ 
\begin{array}{c}
1+2\mathcal{C}_{0,\alpha }\mathcal{U}_{0,\alpha }+2\left( \mathcal{C}%
_{0,\alpha }\mathcal{C}_{0,\beta }-\text{$\frac{1}{2}$}\delta _{\alpha \beta
}\right) \mathcal{U}_{0,\alpha }\mathcal{U}_{0,\beta } \\ 
+\frac{4}{3}\left( \mathcal{C}_{0\alpha }\mathcal{C}_{0\beta }-\text{$\frac{3%
}{2}$}\delta _{\alpha \beta }\right) \mathcal{C}_{0\gamma }\mathcal{U}%
_{0\alpha }\mathcal{U}_{0\beta }\mathcal{U}_{0\gamma }%
\end{array}%
\right] ,
\end{eqnarray}%
where $\vec{\mathcal{U}}_{0}=\frac{\vec{u}}{\left( \frac{2kT_{0}}{m}\right)
^{1/2}}$.\newline
In this case, the quadrature will give for the discrete equilibrium
distribution,

\begin{equation}
f_{i}^{eq}=g_{i}\left( T,\mathcal{C}_{f0,i}^{2}\right) \omega _{i}n\left[ 
\begin{array}{c}
1+2\mathcal{C}_{0,i\alpha }\mathcal{U}_{0,\alpha }+2\left( \mathcal{C}%
_{0,i\alpha }\mathcal{C}_{0,i\beta }-\text{$\frac{1}{2}$}\delta _{\alpha
\beta }\right) \mathcal{U}_{0,\alpha }\mathcal{U}_{0,\beta } \\ 
+\frac{4}{3}\left( \mathcal{C}_{i\alpha }\mathcal{C}_{i\beta }-\text{$\frac{3%
}{2}$}\delta _{\alpha \beta }\right) \mathcal{C}_{i\gamma }\mathcal{U}%
_{0\alpha }\mathcal{U}_{0\beta }\mathcal{U}_{0\gamma }%
\end{array}%
\right] ,  \label{fieq1}
\end{equation}%
where

\begin{equation}
g_{i}\left( T,\mathcal{C}_{f0,i}^{2}\right) =\left( \frac{T_{0}}{T}\right)
^{D/2}\left[ 1+\mathcal{C}_{fo,i}^{2}\Theta \right] ,  \label{tweights1}
\end{equation}%
is a temperature dependent weight. When \ $T>T_{0},$ \ $g_{i}$ reduces the
amount of particles with zero velocity, redistributing them to the kinetic
modes $i$ in accordance with $\mathcal{C}_{f0,i}^{2}$ and inversely when $%
T<T_{0}$. It is well known by LB practitioners that this redistribution is
highly desirable in LB simulation and redistribution rules were empirically
found by some authors (e.g., \cite{alexander}).\newline
Considering a linear approximation to the temperature non-equilibrium,

\begin{equation}
\left( \frac{T_{0}}{T}\right) ^{D/2}\approx 1-\frac{D}{2}\Theta .
\end{equation}

Eq. (\ref{tweights1}) can also be written as

\begin{equation}
g_{i}\left( T,\mathcal{C}_{f0,i}^{2}\right) =1-\frac{D}{2}\Theta +\mathcal{C}%
_{f0,i}^{2}\Theta +O\left( \Theta ^{2}\right) ,
\end{equation}%
and the equilibrium distribution may be written as a sum of two distributions

\begin{equation}
f_{i}^{eq}=f_{i,n}^{eq}+f_{i,T}^{eq},
\end{equation}%
where, dropping-out the third-order term,

\begin{equation}
f_{i,T}^{eq}=\Theta \mathcal{C}_{f0,i}^{2}\omega _{i}n\left[ 1+2\mathcal{C}%
_{0,i\alpha }\mathcal{U}_{0,\alpha }+2\left( \mathcal{C}_{0,i\alpha }%
\mathcal{C}_{0,i\beta }-\text{$\frac{1}{2}$}\delta _{\alpha \beta }\right) 
\mathcal{U}_{0,\alpha }\mathcal{U}_{0,\beta }\right] .
\end{equation}

This equilibrium distribution is related to the thermal distribution
function $g$ in He \textit{et al. }two-distributions model, \cite%
{hechendoolen}. To fit the model into a D2Q9 lattice, He \textit{et al. }%
have, further, replaced $\mathcal{C}_{f0,i}^{2}$ by $\left( \vec{\mathcal{C}}%
_{0,i}-\vec{\mathcal{U}}_{0}\right) ^{2}$, truncating all the terms $O\left( 
\mathcal{C}_{0,i}^{3}\right) $ and higher, after the multiplication. It can
be easily seen that the resulting expression has second-order errors $%
O\left( \Theta \mathcal{U}_{0}\right) $ limiting He \textit{et al}.'s model
to low local speeds.\newline
We have preferred a somewhat different decomposition in Eq. (\ref{dec}),
working with the particles velocity $\vec{c}$ and not with the fluctuation
velocity $\left( \vec{c}-\vec{u}\right) $, making

\begin{equation}
e^{-\frac{c^{2}}{\frac{2kT}{m}}}=\left( e^{-\mathcal{C}_{o}^{2}}\right) ^{%
\frac{T_{0}}{T}},
\end{equation}%
resulting in a temperature dependent weights model (TDW), which equilibrium
distribution is given by

\begin{equation}
f_{i}^{eq}=g_{i}\left( T,\mathcal{C}_{0,i}^{2}\right) \omega _{i}n\left[ 
\begin{array}{c}
1+2\frac{\mathcal{C}_{0,i\alpha }\mathcal{U}_{0,\alpha }}{\frac{T}{T_{0}}}%
+2\left( \mathcal{C}_{0,i\alpha }\mathcal{C}_{0,i\beta }-\text{$\frac{1}{2}$}%
\frac{T}{T_{0}}\delta _{\alpha \beta }\right) \frac{\mathcal{U}_{0,\alpha }%
\mathcal{U}_{0,\beta }}{\left( \frac{T}{T_{0}}\right) ^{2}} \\ 
+\frac{4}{3}\left( \mathcal{C}_{0i\alpha }\mathcal{C}_{0i\beta }-\text{$%
\frac{3}{2}$}\frac{T}{T_{0}}\delta _{\alpha \beta }\right) \frac{\mathcal{C}%
_{0i\gamma }\mathcal{U}_{0\alpha }\mathcal{U}_{0\beta }\mathcal{U}_{0\gamma }%
}{\left( \frac{T}{T_{0}}\right) ^{3}}%
\end{array}%
\right] ,  \label{fieq}
\end{equation}%
where

\begin{equation}
g_{i}\left( T,\mathcal{C}_{0,i}^{2}\right) =1+\left( \mathcal{C}_{0,i}^{2}-%
\frac{D}{2}\right) \Theta .  \label{tweights}
\end{equation}

Since $\frac{T}{T_{0}},\left( \frac{T}{T_{0}}\right) ^{2},...$ in Eq. (\ref%
{fieq}), appears inside the polynomial expansion these terms must be, also,
developed in $\theta $ for preserving consistency in the order of
approximation. After multiplication, an expression for the equilibrium
distribution in 2D problems can be written as

\begin{equation}
f_{i}^{eq}=\omega _{i}n\left\{ 
\begin{array}{c}
1+\left( \mathcal{C}_{0,i}^{2}-1\right) \Theta +2\left[ 1+\left( \mathcal{C}%
_{0,i}^{2}-2\right) \Theta \right] \vec{\mathcal{C}}_{0,i}.\vec{\mathcal{U}}%
_{0} \\ 
+2\left( \vec{\mathcal{C}}_{0,i}\vec{.\mathcal{U}}_{0}\right) ^{2}-\mathcal{U%
}_{0}^{2}+\frac{4}{3}\left( \vec{\mathcal{C}}_{0,i}\vec{.\mathcal{U}}%
_{0}\right) ^{3}-2\left( \vec{\mathcal{C}}_{0,i}.\vec{\mathcal{U}}%
_{0}\right) \mathcal{U}_{0}^{2}%
\end{array}%
\right\} ,  \label{fieqtdw}
\end{equation}%
with errors $O\left( \Theta ^{2}\text{, }\mathcal{U}_{0}^{2}\Theta \right) $%
. A redistribution expression similar to the above one was obtained by Shan
and He, \cite{shanhe}. Large temperature deviations require to consider
additional powers in $\Theta $, in the Taylor expansion, Eq. (\ref{tweights}%
) and to increase the number of discrete velocities in the lattice. This
will be discussed in Section 2.4.3.

\subsection{\textbf{TDV model: allocation rules and collision step}}

Considering the analysis shown in the last section, TDV models appear to be
a promissing alternative for LB simulation of non-isothermal problems, since
no theoretical limitations were identified, related to the thermal
non-equilibrium deviation $\Theta $, as in the TDW models, Eq. (\ref{fieqtdw}%
), where the Taylor series in $\Theta $, Eq. (\ref{tweights}) was truncated
after the first-order term. In present section this model is discussed,
considering that temperature-dependent velocities require to modify the
propagation step in the LB\ simulation scheme.\newline
After quadrature, the third-order approximation to the equilibrium
distribution Eq. (\ref{fiqc1}) can be written in a general form as%
\begin{equation}
\frac{f_{k}^{eq}}{\omega _{k}n}=%
\begin{array}{c}
1+2a^{2}\frac{c_{k\alpha }^{\ast }u_{\alpha }^{\ast }}{\frac{T}{T_{0}}}%
+2a^{4}\left( c_{k\alpha }^{\ast }c_{k\beta }^{\ast }-\frac{1}{2a^{2}}\frac{T%
}{T_{0}}\delta _{\alpha \beta }\right) \frac{u_{\alpha }^{\ast }u_{\beta
}^{\ast }}{\left( \frac{T}{T_{0}}\right) ^{2}} \\ 
+\frac{4}{3}a^{6}\left( c_{k\alpha }^{\ast }c_{k\beta }^{\ast }-\frac{3}{%
4a^{2}}\frac{T}{T_{0}}\delta _{\alpha \beta }\right) \frac{c_{k\gamma
}^{\ast }u_{\alpha }^{\ast }u_{\beta }^{\ast }u_{\gamma }^{\ast }}{\left( 
\frac{T}{T_{0}}\right) ^{3}}%
\end{array}%
,  \label{fieqtdv}
\end{equation}%
where $a$ is a lattice constant related to the lattice symmetry. The $\omega
_{k}$ are the lattice-weights, defining the inner product

\begin{equation}
f.g=\sum_{k}\omega _{k}f_{k}g_{k}.
\end{equation}

In general, the discrete velocities set generated by quadrature will be not
space-filling requiring simulation strategies similar to the one proposed by
Pavlo \textit{et al.}, \cite{vahala}. In space-filling lattices such as,
e.g., the D2Q13H, the equilibrium distribution will keep the form given by
Eq. (\ref{fieqtdv}) but the weights $\omega _{k}$ must be, empirically
chosen for ensuring isotropy of even-parity rank velocity tensors up to the 6%
$^{th}$-rank.\newline
Particle velocities are given by

\begin{equation}
\vec{c}_{k}=\frac{h}{\delta }\vec{c}_{k}^{\ast }.
\end{equation}%
where the lattice dimension $h$ is given by

\begin{equation}
\frac{h}{\delta }=a\left( \frac{2kT_{0}}{m}\right) ^{1/2},
\end{equation}%
and

\begin{equation}
\vec{c}_{k}^{\ast }=\sqrt{\frac{T}{T_{0}}}\vec{e}_{k},
\end{equation}%
where $\left\{ \vec{e}_{k},k=0,...,b\right\} $ are the lattice unity-vectors.

In present scheme, macroscopic velocity $\vec{u}^{\ast }=$ $\vec{u}/\left( 
\frac{h}{\delta }\right) $, is calculated as $n\vec{u}^{\ast }=\sum_{k}f_{k}%
\vec{c}_{k}^{\ast }$ where $n=\sum_{k}f_{k}$. Temperature is found to be
such that 
\begin{equation}
n\frac{T}{T_{0}}=a^{2}\sum_{k}f_{k}\left( \vec{c}_{k}^{\ast }-\vec{u}^{\ast
}\right) ^{2}.  \label{TD}
\end{equation}

After discretization, the second-order model reads, in two-dimensions reads

\begin{eqnarray}
&&\widetilde{f_{k}}\left( \vec{x}+\vec{c}_{k}^{\ast }\delta ,t+\delta \right)
\notag \\
=\widetilde{f_{k}}\left( \vec{x},t\right) &&+\frac{\delta }{\tau _{_{3}}+%
\frac{1}{2}\delta }\left( f_{k}^{eq}-\widetilde{f}_{k}\right) -\frac{\delta 
}{\tau _{_{3}}+\frac{1}{2}\delta }\frac{1}{\left( 1+\frac{\delta }{2\tau
_{_{3}}}+\frac{\delta }{2\tau _{_{\mu }}}\right) }\text{ \ }  \notag \\
\times 6a^{2} &&\frac{\tau _{_{3}}}{\tau _{_{\mu }}}\frac{1}{\left( \frac{T}{%
T_{0}}\right) ^{2}}f_{k}^{eq}\left[ 
\begin{array}{c}
\widetilde{{\LARGE \tau }}_{xx}^{\ast }\left( c_{fk,x}^{\ast 2}-\frac{1}{%
2a^{2}}\frac{T}{T_{0}}\right) + \\ 
\widetilde{{\LARGE \tau }}_{yy}^{\ast }\left( c_{fk,y}^{\ast 2}-\frac{1}{%
2a^{2}}\frac{T}{T_{0}}\right) +2\widetilde{{\LARGE \tau }}_{xy}^{\ast
}c_{fk,x}^{\ast }c_{fk,y}^{\ast }%
\end{array}%
\right] ,
\end{eqnarray}%
where $\tau _{_{3}}=1/\gamma _{_{3}}$ and $\tau _{_{\mu }}=1/\lambda _{_{\mu
}}$.

The dimensionless viscous stress tensor is calculated as,

\begin{equation}
{\LARGE \tau }_{\alpha \beta }^{\ast }=\frac{{\LARGE \tau }_{\alpha \beta }}{%
2a^{2}nkT_{0}}=\frac{1}{n}\sum_{k}f_{k}^{neq}c_{k\alpha }^{\ast }c_{k\beta
}^{\ast },
\end{equation}%
and the heat flux, as

\begin{equation}
q_{\gamma }^{\ast }=\frac{q_{\gamma }}{a^{2}kT_{0}\frac{h}{\delta }}%
=\sum_{k}f_{k}^{neq}c_{fk\gamma }^{\ast }c_{fk}^{\ast 2}.
\end{equation}

The modified fluxes $\widetilde{{\LARGE \tau }}_{\alpha \beta }^{\ast }$ and 
$\widetilde{q}_{\gamma }^{\ast }$ needed in the simulation have similar
expressions in terms of $\widetilde{f}_{k}^{neq}$.

The particles that are present at site $\vec{x}$ after the collision step
have a velocity $\vec{c}_{i}^{\ast }=\sqrt{\frac{T}{T_{0}}}\vec{e}_{i}$.
When $\sqrt{\frac{T}{T_{0}}}\leq 1$, the kinetic energy these particles have
is just enough to enable them to jump to some intermediate position between $%
\vec{x}$ and $\vec{x}+\vec{e}_{i}$. In present \textit{discrete }model these
particles are redistributed to the next contiguous sites $\vec{x}$ and $\vec{%
x}+\vec{e}_{i}$ preserving the mass, kinetic energy and the momentum of the
original particles packet $\widetilde{f}_{i}\left( \vec{x},t\right) $. This
is performed by using a lever\'s rule, making $\widetilde{f}_{i,0}\left( 
\vec{x},t+\delta \right) =\widetilde{f}_{i}\left( \vec{x},t\right) \left( 1-%
\frac{T}{T_{0}}\right) $ and $\widetilde{f}_{i,-1}\left( \vec{x}+\vec{e}%
_{i},t+\delta \right) =\widetilde{f}_{i}\left( \vec{x},t\right) \left( \frac{%
T}{T_{0}}\right) $, and by imposing to both amounts $\widetilde{f}_{i,0}$
and $\widetilde{f}_{i,-1}$ the same \textit{non-integer} velocity $\vec{c}%
_{i}^{\ast }$ related to the particles-packet before the propagation. In
this way model particles are \textit{velocity-memory} particles in the
propagation step. In the notation $\widetilde{f}_{ij}$, the index $i$ is
related to the site direction and $j$ means from what site these particles
were come: $j=0$ means that the particles were come from the same site, $%
j=-1 $, from the site $\vec{x}-\vec{e}_{i}$, $\ j=-2$, from the site $\vec{x}%
-2\vec{e}_{i}$ and so on.\newline
In this manner, the particles propagated along the direction $i$ are
redistributed among the departure site and the next-one contiguous site, but
have their velocity unaltered, after the propagation. These rules preserve,
locally, mass, momentum and energy and non-physical convection is avoided.%
\newline
When \ $1$ $\leq \sqrt{\frac{T}{T_{0}}}\leq 2$ , particles will be
redistributed following

\begin{equation}
\widetilde{f}_{i,-1}\left( \vec{x}+\vec{e}_{i},t+\delta \right) =\widetilde{f%
}_{i}\left( \vec{x},t\right) \frac{1}{3}\left( 4-\frac{T}{T_{0}}\right) ,
\end{equation}
and

\begin{equation}
\widetilde{f}_{i,-2}\left( \vec{x}+2\vec{e}_{i},t+\delta \right) =\widetilde{%
f}_{i}\left( \vec{x},t\right) \frac{1}{3}\left( \frac{T}{T_{0}}-1\right) .
\end{equation}

Temperature ratios $\sqrt{\frac{T}{T_{0}}}$ \ greater than 2 are not
expected in present work, since they would represent very strong thermal
non-equilibrium.\newline
After propagation, at time step $t$, all the distributions $\widetilde{f}%
_{i,j}\left( \vec{x},t\right) $ are known in each site $\vec{x}$. The
amounts $\widetilde{f}_{i,j}\left( \vec{x},t\right) $ enable to calculate
the \textit{equilibrium moments} $n(\vec{x},t)$, $u^{\ast }(\vec{x},t)$ and $%
T(\vec{x},t)$\textit{\ }and the equilibrium distribution, Eq. (\ref{fieqtdv}%
), with $k=(i,j)$.\newline
The non-equilibrium distributions are given by $\widetilde{f}_{i,j}^{neq}=%
\widetilde{f}_{i,j}-f_{ij}^{eq}$ and the viscous stress tensor can be thus
calculated. In the collision step, collision will give a \textit{single}
distribution $\widetilde{f_{i}}^{\prime }\left( \vec{x},t\right) $ from the
several $j$\ amounts $\widetilde{f}_{i,j}$ in the direction $i$ of the site $%
\vec{x}$ at time $t$,%
\begin{equation}
\widetilde{f_{i}}^{\prime }\left( \vec{x},t\right) =\sum_{j}\left\{ 
\begin{array}{c}
\widetilde{f}_{i,j}+\frac{\delta }{\tau _{_{N}}+\frac{1}{2}\delta }\left(
f_{ij}^{eq}-\widetilde{f}_{i,j}\right) -\frac{\delta }{\tau _{_{N}}+\frac{1}{%
2}\delta }\frac{6a^{2}\frac{\tau _{3}}{\tau _{\mu }}}{\left( 1+\frac{\delta 
}{2\tau _{_{N}}}+\frac{\delta }{2\tau _{\mu }}\right) }\times \\ 
\text{ }\frac{1}{\left( \frac{T}{T_{0}}\right) ^{2}}f_{ij}^{eq}\left[ 
\begin{array}{c}
\widetilde{{\LARGE \tau }}_{xx}^{\ast }\left( c_{f,ij,x}^{\ast 2}-\frac{1}{%
2a^{2}}\frac{T}{T_{0}}\right) +\widetilde{{\LARGE \tau }}_{yy}^{\ast }\left(
c_{f,ij,y}^{\ast 2}-\frac{1}{2a^{2}}\frac{T}{T_{0}}\right) \\ 
+2\widetilde{{\LARGE \tau }}_{xy}^{\ast }c_{f,ij,x}^{\ast }c_{f,ij,y}^{\ast }%
\end{array}%
\right]%
\end{array}%
\right\} .
\end{equation}

These particles will move to the next sites with the site velocity $\vec{%
c^{\ast }}_{i}=\sqrt{\frac{T}{T_{0}}}\vec{e}_{i}$, where $T=T(\vec{x},t)$ is
the site temperature, calculated using Eq.( \ref{TD}), just after the
propagation step and with $k$ replaced by $(i,j)$.\newline
In the propagation step, a fraction of the low-speed particle-packets, in
sites where $T$ is small, will remain in its departure site. The overall
effect is to increase the number-density of particles in the sites where the
temperature is small and to decrease it in the sites where the temperature
is high, emulating the temperature dependence of the MB distribution. In
this way, the equilibrium distribution is temperature dependent even when
the macroscopic velocity $\vec{u}$ is zero, i.e., when $f_{i}^{eq}=\omega
_{i}n$, because $n$ is temperature dependent. Also, the temperature
derivative of the equilibrium distribution will vary inversely with the
temperature, similarly to its continuous counterpart. This is an important
condition for retrieving the temperature derivative term in the second
member of Eq.(\ref{df0}).\newline
Nevertheless, the temperature dependence of $n$ is related to the particles
allocation rules used and further theoretical analysis of presently proposed
procedure leading to a correct Chapman-Enskog analysis of TDV model is,
currently, under investigation. \textbf{\ }

\subsection{The nine bits lattice}

In this section, the TDW and TDV models are explicitly written for the D2Q9
lattice and their feasibility for simulating isothermal and non-isothermal
problems in this lattice is discussed.

\subsubsection{\protect\bigskip TDW model}

The D2Q9 lattice is not suitable for thermal problems since: i) 6$^{th}$
rank tensors are not isotropic, ii) energy transfer $q_{\alpha \beta \gamma
} $ is not correctly described in this lattice and iii) the
lattice-dimensionality is too small. However, the use of thermal models in
the D2Q9 lattice can, possibly, be shown to be a promising alternative for
simulating \textit{near}-\textit{isothermal} problems, when the main
interest is to increase numerical stability, with respect to athermal
simulation. In fact, in its kinetic theory concept, temperature is varying
from site to site, with the local velocity $\vec{u}$. When this variation is
not considered, i.e., when the growing of temperature gradients are not
annihilated by heat flow, they can, possibly, become instability sources in
the numerical scheme. Actually, in contrast with classical CFD simulation of
isothermal Navier-Stokes equations, lattice-Boltzmann simulation is never 
\textit{isothermal} and the consideration of heat flow should be helpful in
the simulations.\newline
For the D2Q9 lattice, $r_{p_{m}}(b)=5=2X3-1$. Gauss-Hermite quadrature
gives, \cite{heluo}

\begin{equation}
\vec{\mathcal{C}}_{0,i}=\sqrt{\frac{3}{2}}\vec{e}_{i}
\end{equation}

where $\vec{e}_{i}$ = $(0,0),(1,0),(0,1),(-1,0),(0,-1),(1,1),(-1,1),$ $%
(-1,-1),(1,-1).$ The weights $\omega _{i}$ are found to be $\omega
_{_{0}}=4/9,\omega _{i}=1/9,i=1,...,4$ \ for the main axes and $\omega
_{i}=1/36,i=5,...,8$, for the diagonals.\newline
Using a dimensionless macroscopic velocity $\vec{u}^{\ast }=\frac{\vec{u}}{%
\frac{h}{\delta }}$ and dropping-out the third-order term, since, as
commented above, the exact description of energy transfer is \textit{out of
thought} in this lattice, the equilibrium distribution \ in the TDW model
will read

\begin{equation}
f_{i}^{eq}=\omega _{i}n\left[ 
\begin{array}{c}
1+\frac{3}{2}\left( e_{i}^{2}-\frac{2}{3}\right) \Theta +3\left[ 1+\frac{3}{2%
}\left( e_{i}^{2}-\frac{4}{3}\right) \Theta \right] \vec{e}_{i}.\vec{u}%
^{\ast } \\ 
+\frac{9}{2}\left( \vec{e}_{i}.\vec{u}^{\ast }\right) ^{2}-\frac{3}{2}\left(
u^{\ast }\right) ^{2}%
\end{array}%
\right] .  \label{eq}
\end{equation}

It can be easily seen that the equilibrium distribution given by Eq. (\ref%
{eq}) satisfies

\begin{equation}
n=\sum_{i}f_{i}^{eq},  \label{R1}
\end{equation}

\begin{equation}
n\vec{u}=\sum_{i}f_{i}^{eq}\vec{c}_{i},  \label{R2}
\end{equation}

\begin{equation}
P=nkT.  \label{R3}
\end{equation}
\ 

The internal energy

\begin{equation}
E=\sum_{i}f_{i}^{eq}\frac{1}{2}m(\vec{c}_{i}-\vec{u})^{2},  \label{R4}
\end{equation}%
the momentum flux

\begin{equation}
\Pi _{\alpha \beta }^{eq}=\sum_{i}f_{i}^{eq}mc_{i\alpha }c_{i\beta ,}
\label{R5}
\end{equation}

and the heat flux

\begin{equation}
q_{\alpha }^{eq}=\sum_{i}f_{i}^{eq}\frac{1}{2}m(\vec{c}_{i}-\vec{u}%
)^{2}\left( c_{i\alpha }-u_{\alpha }\right) ,  \label{R6}
\end{equation}%
are also retrieved as, respectively, $\frac{D}{2}nkT$, $P\delta _{\alpha
\beta }+\rho u_{\alpha }u_{\beta }$ and $0$, with errors $O\left( \mathcal{U}%
_{0}^{2}\Theta ,\Theta ^{2}\right) $.

\subsubsection{TDV model}

All the restrictions given by Eqs.(\ref{R1}-\ref{R3} and \ref{R4}-\ref{R6})
are satisfied when the equilibrium distribution is given by Eq. (\ref%
{fieqtdv}), dropping-out the third-order term, i.e.,

\begin{equation}
f_{k}^{eq}=\omega _{k}n\left[ 1+3\frac{c_{k\alpha }^{\ast }u_{\alpha }^{\ast
}}{\frac{T}{T_{0}}}+\frac{9}{2}\left( c_{k\alpha }^{\ast }c_{k\beta }^{\ast
}-\frac{1}{3}\frac{T}{T_{0}}\delta _{\alpha \beta }\right) \frac{u_{\alpha
}^{\ast }u_{\beta }^{\ast }}{\left( \frac{T}{T_{0}}\right) ^{2}}\right] .
\label{eqtdv}
\end{equation}

Particles redistribution in TDV model should increase numerical stability
with respect to the athermal simulation method, since, in this last scheme,
particle allocation is not temperature dependent and sites with higher $T$
will rest with an excess amount in the number of particles, contributing to
the enhancement of non-physical source-terms.\newline
In Pavlo \textit{et al. }, \cite{vahala}, TDV model, the use of an octagonal
lattice, instead of the D2Q9 lattice, was considered for ensuring \ isotropy
for the 6$^{th}$-rank tensors. A second-order interpolation scheme is used
for allocating the particles among adjacent sites in the propagation step,
following a procedure that is similar to the one exposed in Section 2.3, but
the model particles are not velocity-memory particles and the D2Q9 $\vec{e}%
_{i}$ lattice-velocities are attributed to the redistributed packets. This
is performed preserving mass, momentum and energy. After the propagation,
the number density, the momentum and the kinetic energy are calculated in
each site. Previously to the collision step, the distributions $f_{i}$ are
renormalized to account for the octagonal lattice velocities, satisfying the
new local moment constraints. Successive transitions between the D2Q9 and
the octagonal lattice can affect the isotropy properties of the 6$^{th}$%
-rank tensors and, in our opinion, this is the only question that remains to
be answered, since numerical viscosity effects can be avoided by reducing
the spatial scale. In this manner, the problem is, nearly, the same as
above, because the correct description of the heat flow vector cannot be
assured in this model.

\subsubsection{Increasing the number of discrete velocities}

After the very promising results of He and Luo's work, \cite{heluo}, when
the D2Q9, D2Q7 and D3Q27 lattices were found by exact quadrature, it is,
now, apparent that the integer multiplicative factors that are required by
regular-lattice velocity components are a too strong restriction: for their
exact quadrature, Maxwellian-shaped curves require the roots $c_{i\alpha }$
to be progressively close when their absolute values increase.\newline
For some regular lattices, the weights $\omega _{i}$ can be found to give
even-parity isotropic tensors. In this manner, isotropy up to the 6$^{th}$%
-rank tensors is assured for the D2Q13H lattice by choosing $\omega
_{0}=132/300$, for the zero velocity, $\omega _{1}=27/300$ for the \ first 6
velocity vectors, with $\left\vert \vec{e}_{i}\right\vert =1$ and $\omega
_{2}=1/300$ for the next 6 velocity vectors, with $\left\vert \vec{e}%
_{i}\right\vert =\sqrt{3}$. In this lattice $h/\delta =\sqrt{\frac{10kT_{0}}{%
3m}\text{.}}$\newline
Simulation of non-isothermal problems is possible in this lattice by using
third-order approximations to the MB equilibrium distribution, derived,
alternatively, from Eq. (\ref{fieqtdw}) in the TDW model or from Eq. (\ref%
{fieqtdv}) in the TDV model.\newline
In TDW models, improving the approximation, Eq. (\ref{fieqtdw}), to the MB
equilibrium distribution, considering $\Theta ^{2}$ powers in Eq. (\ref%
{tweights3}), also requires to take 4$^{th}$-degree powers of $\ \mathcal{C}%
_{0}$ into account and, consequently, to consider 4$^{th}$- degree
polynomial expansions in $\mathcal{U}_{0}$ for the equilibrium distribution.
This has been performed for a D2Q21S lattice, choosing the weights $\omega
_{i}$ for giving even-parity isotropic tensors up to the 6$^{th}$-rank and
for retrieving the equilibrium conditions, Eqs.(\ref{R1}-\ref{R3} and \ref%
{R4}-\ref{R6}), with errors $O\left( \mathcal{U}_{0}^{3}\Theta ,\Theta ^{3},%
\mathcal{U}_{0}^{5},\mathcal{U}_{0}\Theta ^{2},...\right) $. The D2Q21S is
composed by the zero velocity and by the superposition of five square
lattices with a $45^{0}$ shift between each two sequential lattices, with
speeds $0$, $1$, $\sqrt{2}$, $2$, $2\sqrt{2}$ and $3$. The equilibrium
distribution was found as

\begin{equation}
f_{i}^{eq}=\omega _{i}n\left[ 
\begin{array}{c}
1+\left( e_{i}^{2}-1\right) \Theta +\left( 1-2e_{i}^{2}+e_{i}^{4}/2\right)
\Theta ^{2}+2\left[ 1+\left( e_{i}^{2}-2\right) \Theta \right] \vec{e_{i}}.%
\vec{u}^{\ast } \\ 
+2\left( \vec{e_{i}}.\vec{u}^{\ast }\right) ^{2}\left( 1+\left(
e_{i}^{2}-3\right) \Theta \right) -\left( u^{\ast }\right) ^{2}\left(
1+\left( e_{i}^{2}-2\right) \Theta \right) \\ 
+\frac{4}{3}\left( \vec{e}_{i}.\vec{u}^{\ast }\right) ^{3}-2\left( \vec{e}%
_{i}.\vec{u}^{\ast }\right) \left( u^{\ast }\right) ^{2}+\frac{2}{3}\left( 
\vec{e}_{i}.\vec{u}^{\ast }\right) ^{4} \\ 
-2\left( \vec{e}_{i}.\vec{u}^{\ast }\right) ^{2}\left( u^{\ast }\right) ^{2}+%
\frac{1}{2}\left( u^{\ast }\right) ^{4}%
\end{array}%
\right] .
\end{equation}

with weights $\omega _{i}=\left\{ \frac{379}{1152},\frac{41}{384},\frac{5}{96%
},\frac{31}{3840},\frac{1}{1536},\frac{1}{5760}\right\} $.

\section{Conclusion}

As a kinetic method, LBM is based on a discrete and finite approximation to
the continuous Boltzmann equation giving an approximated solution to the
particles distribution function. In this manner, all the macroscopic moments
of interest will be affected \ by the accuracy of the modelled LBE when
compared to its full continuous counterpart.\newline
MRT increasing accuracy models to the collision term $\Omega $ were derived
collapsing the higher-order terms, related to the relaxation of the
higher-order moments, into a single non-equilibrium term, minimizing the
truncation effects on the fine structure of the collision operator spectrum.%
\newline
Thus, in contrast with the moments method, only the hydrodynamic moments
that are required to be correctly described, are considered in the present
models. This strategy avoids the use of dispersion relations for buffering
undesirable effects of the high-frequency, non-hydrodynamic moments, but
claims to increase the lattice dimensionality.\newline
Time discretization requires implicit or modified explicit numerical schemes
for avoiding $\mathcal{O}(\delta ^{2})$ time step errors and non-physical
source terms in the internal energy balance equation.\newline
For the discretization of the velocity space, the requirements for the
Boltzmann equation quadrature to be exact establishes the minimal discrete
velocity set that is needed for the lattice, in accordance with the order of
approximation for the kinetic model that is intended to be used.
Nevertheless, these sets do not generate regular lattices when multi-speed
models appropriated for thermal problems are considered.\newline
When performing the quadrature, it was shown that the integrating variable
has an important role in defining the equilibrium distribution and the
lattice-Boltzmann model, leading, alternatively, to temperature dependent
velocities (TDV) and to temperature dependent weights (TDW)
lattice-Boltzmann models.\newline
In TDV models no theoretical limitations were identified, related to the
thermal non-equilibrium deviation $\Theta $, as in the TDW models.\newline
Finding the macroscopic behavior of TDV models through a rigorous
Chapman-Enskog analysis is difficult due to the implicit number density of
particles dependence on the temperature and is, still, in progress.

\textbf{Acknowledgements}

Authors are greatly indebted to ANP (Brazilian Petroleum Agengy), CNPq
(Brazilian Research Council), Finep (Brazilian Agency for Research and
Projects) and Petrobras (Brazilian Petroleum Company) for the financial
support.


\begin{thebibliography}{99}
\bibitem{lallemandluo} Lallemand, P. and Luo, L. S.: Theory of the lattice
Boltzmann method: Acoustic and thermal properties in two and three
dimensions, Phys. Rev. E, 68(3):036706, 2003.

\bibitem{bgk} Bhatnagar, P.L., Gross, E.P. and Krook, M.: A model for
collisional processes in gases: small amplitude processes in charged and
neutral one-component system, Phys. Rev. 94, 511, 1954.

\bibitem{qian} Qian, Y.H., D'Humi\`{e}res, D. and Lallemand, P.: Lattice BGK
models for Navier-Stokes equation. Europhys. Lett., 17(6): 479-484, 1992.

\bibitem{chen} Chen, H., Chen, S. and Mathaeus, W.: Recovery of the
Navier-Stokes equation using a lattice-gas Boltzmann method. Phys. Rev. A,
45(8): 5339-5342, 1992.

\bibitem{alexander} Alexander, F.J., Chen, S., Sterling, J.D.: Lattice
Boltzmann themohydrodynamics, Phys. Rev. E 47, R2249, 1993.

\bibitem{mcnamara} McNamara, G. and Alder, B. J.: Analysis of the lattice
Boltzmann treatment of hydrohynamics, Physica A, 194: 218-228, 1993.

\bibitem{ychen} Chen, Y., Ohashi, H. and Akyama, M.: Thermal lattice
Bhatnagar-GrossKrook model without non-linear deviations in macrodynamic
equations, Phys. Rev. E 50, 2776-2783, 1994.

\bibitem{hechendoolen} He, X., Chen, S. Doolen, G.D.: A novel thermal model
for the lattice Boltzmann method in incompressible limit, J. Comp. Phys. 
\textbf{146}, 282-300, 1998.

\bibitem{dhumieres} d'Humi\`{e}res, D.: Generalized Lattice Boltzmann
Equations, in : Rarefied Gas Dynamics: Theory and Simulations, Prog.
Astronau. Aeronaut. \textbf{159}, 450-458, 1992.

\bibitem{dhumieresetal} d'Humi\`{e}res, D., Bouzidi, M. and Lallemand, P.:
Thirteen-velocity three-dimensional lattice Boltzmann model, Phys. Rev. E 
\textbf{63}: 66702, 2001.

\bibitem{heluo} He, X., Luo, L.S.: Theory of the lattice-Boltzmann method:
>From the Boltzmann equation to the lattice-Boltzmann equation, Phys. Rev. E
56, 6811-6817, 1997.

\bibitem{vahala} Pavlo, P., Vahala, G. , Vahala, L.: Preliminary results in
the use of energy-dependent octagonal lattices for thermal lattice Boltzmann
simulations,J. Stat. Phys. 107, 499-519, 2002.

\bibitem{krafk} Toelke, J., Krafczyk, M., Schulz, M., Rank, E., Berrios, R.:
Implicit discretization and non-uniform mesh refinement approaches for FD
discretizations of LBGK Models, Int. J. of Mod. Physics C 9, 1143-1157, 1998.

\bibitem{sanka} Sankaranarayanan, K., Shan, X., Kevrekidis, I.G.,
Sundaresan, S.: Analysis of drag and virtual mass forces in bubbly
suspensions using an implicit formulation of the lattice Boltzmann method,
J. Fluid. Mech \textbf{452}, 61-96, 2002.

\bibitem{cercignani} Cercignani, C.: Mathematical Methods in Kinetic Theory,
First edition, Macmillan, London, 1969

\bibitem{philippi} Philippi, P. C. and Brun, R.: Kinetic modeling of
polyatomic gas mixtures, Physica A, 105(1-2): 147-168 , 1981.

\bibitem{grad} Grad, H.: Principles of the Kinetic Theory of Gases, in :
Handbuch der Physik 12, Springer, New York, 205-294, 1958.

\bibitem{wolfram} Wolfram, S.: Cellular automation fluids 1: basic theory,
J. Stat. Phys. 45, 471-526, 1986.

\bibitem{shanhe} Shan, X. and He, X.: Discretization of the velocity space
in the solution of the Boltzmann equation, Phys. Rev. Lett. 80,65-68, 1998.
\end{thebibliography}
\end{document}